\documentclass[11 pt]{article}
\usepackage[centertags]{amsmath}
\usepackage{amsfonts}
\usepackage{amssymb}
\usepackage{amsthm}
\usepackage{amsrefs}
\usepackage{hyperref}
\usepackage{newlfont}
\usepackage{xy}
\xyoption{all}
\usepackage{color}
\usepackage{epsfig}
\usepackage{fancyhdr}
\usepackage{graphicx}
\usepackage{float}
\usepackage{lscape}
\usepackage{pdflscape}

\newtheorem{theorem}{Theorem}[section]
\newtheorem{prop}[theorem]{Proposition}

\newcommand{\RR}{\mathbb{R}}

\renewcommand{\th}{\text{th}}
\newcommand{\e}{\mathrm{e}}

\newcommand{\p}{\partial}

\newcommand{\cf}{\bar{f}}
\newcommand{\cp}{\bar{p}}

\newcommand{\al}{\alpha}

\newcommand{\si}{\sigma}

\newcommand{\om}{\omega}

\newcommand{\Up}{\Upsilon}
\newcommand{\ka}{\kappa}

\newcommand{\glap}[1]{\Box_{#1}}

\newcommand{\lap}{\Delta}

\newcommand{\Schrodinger}{Schr\"{o}dinger }

\renewcommand{\Re}{\operatorname{Re}}

\newcommand{\pnth}[1]{\left( #1 \right)}

\newcommand{\abs}[1]{\left| #1 \right|}

\newcommand{\brce}[1]{\left\{ #1 \right\}}
\newcommand{\brkt}[1]{\left[ #1 \right]}

\title{Spherically Symmetric Static States of Wave Dark Matter}
\author{Alan R. Parry\footnote{Email: alan.parry@uconn.edu} \\ {\em Department of Mathematics, University of Connecticut,} \\ {\em 196 Auditorium Road, Unit 3009, Storrs, CT 06269-3009, USA}}
\date{\today}

\begin{document}

\maketitle

\begin{abstract}
  In this paper, we present two observations about static spherically symmetric solutions of the Einstein-Klein-Gordon equations.  The first is a comment extending the well-known result of the existence of static states (i.e.\ standing wave solutions) of the Einstein-Klein-Gordon equations.  The second more important observation shows that, in the low field limit, the mass profiles of these static states lie along hyperbolas of constant $\Up$, the fundamental constant of the Einstein-Klein-Gordon equations.
\end{abstract}

\section{Introduction}

The study of scalar fields in general relativity satisfying the Einstein-Klein-Gordon equations has been common in the literature for a very long time (see \cites{Feinblum68, Kaup68} among others).  In the last few decades, these solutions have been considered as a possible candidate for dark matter \cites{Lee09, Matos09, MSBS, Seidel90, Seidel98, Bernal08, Sin94, Mielke03, Sharma08, Ji94, Lee92, Lee96, Guzman01,Guzman00}.  Often these solutions are referred to as boson stars or scalar field dark matter \cites{Bizon00, Pugliese13, Jetzer92B, Kusmart91, Schunck98,Lee09, Matos09, MSBS, Seidel90, Seidel98, Bernal08, Sin94, Mielke03, Sharma08, Ji94, Lee92, Lee96, Guzman01,Guzman00}.  Recently, Bray has added to this discussion compelling geometrical reasons to advocate scalar fields as a candidate for dark matter \cites{Bray10,Bray12}.  In his work, due to the fact that the Klein-Gordon equation is a wave-like equation, he referred to these solutions as wave dark matter and we have adopted that name here as well.  Specifically, wave dark matter refers to any solution of the Einstein-Klein-Gordon equations whether static, spherically symmetric, or otherwise.

In this paper, we make a few useful remarks on the properties of static spherically symmetric solutions of wave dark matter.  The solutions we consider here are usually referred to as static states or boson stars (see \cites{Seidel98, Seidel90, MSBS, Bizon00} among many others).

\section{Spherically Symmetric Static States}

To describe the Einstein-Klein-Gordon equations in spherical symmetry, we use the setup discussed in the survey article by the author \cite{Parry12-1}.  We include here for convenience only the equations from \cite{Parry12-1} and refer the reader to the detailed discussion about them there.  Consider the Einstein-Klein-Gordon system
\begin{subequations}\label{EKG}
    \begin{align}
      \label{EKG-1} G &= 8\pi \mu_{0}\pnth{\frac{df \otimes d\cf + d\cf \otimes df}{\Up^{2}} - \pnth{\frac{\abs{df}^{2}}{\Up^{2}} + \abs{f}^{2}}g}, \\
      \label{EKG-2} \glap{g}f &= \Up^2 f,
    \end{align}
\end{subequations}
where $f$ is a complex-valued scalar field and $\mu_{0}$ and $\Up$ are positive real constants. 
We work in spherical symmetry with the metric in polar-areal coordinates written in the form
\begin{equation}\label{metric}
  g = -\e^{2V(t,r)}\, dt^{2} + \pnth{1 - \frac{2M(t,r)}{r}}^{-1}\, dr^{2} + r^{2}\, d\si^{2}
\end{equation}
for real valued functions $V$ and $M$.  In this case, the Einstein-Klein-Gordon equations reduce to the following set of ODEs \cite{Parry12-1}.
\begin{subequations} \label{NCpde}
  \begin{align}
    \label{NCpde1a} M_{r} &= 4\pi r^{2}\mu_{0}\pnth{\abs{f}^{2} + \pnth{1 - \frac{2M}{r}}\frac{\abs{f_{r}}^{2} + \abs{p}^{2}}{\Up^{2}}}, \\
    \label{NCpde2a} V_{r} &= \pnth{1 - \frac{2M}{r}}^{-1}\pnth{\frac{M}{r^{2}} - 4\pi r\mu_{0}\pnth{\abs{f}^{2} - \pnth{1 - \frac{2M}{r}}\frac{\abs{f_{r}}^{2} + \abs{p}^{2}}{\Up^{2}}}}, \displaybreak[0]\\
    \label{NCpde3a} f_{t} &= p\e^{V}\sqrt{1 - \frac{2M}{r}}, \\
    \label{NCpde4a} p_{t} &= \e^{V}\pnth{-\Up^{2}f\pnth{1 - \frac{2M}{r}}^{-1/2} + \frac{2f_{r}}{r}\sqrt{1 - \frac{2M}{r}}} + \p_{r}\pnth{\e^{V}f_{r}\sqrt{1 - \frac{2M}{r}}}.
  \end{align}
\end{subequations}
Note also the following equation 
which is automatically satisfied by satisfying the equations above \cite{Parry12-1}.
\begin{equation}
    \label{NCpde5a} M_{t} = \frac{8\pi r^{2}\mu_{0}\e^{V}}{\Up^{2}}\pnth{1 - \frac{2M}{r}}^{3/2}\Re(f_{r}\cp).
\end{equation}

Next, we employ the common ansatz that $f$ is of the form
\begin{equation}\label{sfstaticstate}
  f(t,r) = \e^{i\om t}F(r),
\end{equation}
with $\om \in \RR$.  Usually, when this ansatz is used, $F$ is assumed to be real-valued (see \cite{Kaup68,Bizon00,MSBS} among others).  Bizon and Wasserman \cite{Bizon00} gave a description of the solutions produced in this case called static states.  But one might ask, why assume $F$ is real-valued?  Our first comment concerns this question.  If $F$ is complex-valued and of special type, then it turns out that the only solutions are still those found in \cite{Bizon00}.  The following proposition can also be found in the author's dissertation with a more detailed proof \cite{ParryPhD}.  We only present the most interesting part of the proof here.

\begin{prop}\label{staticprop}
Let $(N,g)$ be a spherically symmetric asymptotically Schwarzschild spacetime that satisfies the Einstein-Klein-Gordon equations (\ref{EKG}) for a scalar field of the form in (\ref{sfstaticstate}).  Additionally, assume that $F(r) = h(r)\e^{i a(r)}$ for smooth real-valued functions $h$ and $a$, where $h$ has only isolated zeros, if any.  Then $(N,g)$ is static if and only if $a(r)$ is constant.
\end{prop}

\begin{proof}  If $a(r)=0$, then the fact that $(N,g)$ is static is shown in \cite{Bizon00}.  It is shown that $(N,g)$ is static if and only if $M_{t}\equiv 0$ in the extended proof of this proposition in \cite{ParryPhD}.  By equation (\ref{NCpde5a}), \mbox{$M_{t} \equiv 0$} if and only if $\Re(f_{r}\cp) \equiv 0$.  A short computation yields that this is true if and only if \mbox{$\Re(-i \om F'(r) \overline{F(r)}) \equiv 0$} or equivalently that $F'(r)\overline{F(r)}$ is real valued.  By assumption, $F(r)$ can be written as
  \begin{equation}
    F(r) = h(r)\e^{i a(r)}
  \end{equation}
  for smooth real-valued functions $h$ and $a$.  
  Then we have that
  \begin{equation}
    F'(r)\overline{F(r)} = h'(r)h(r) + i h(r)^2 a'(r).
  \end{equation}
  Since $h$ and $a$ are both real-valued, we see that $F'(r)\overline{F(r)}$ is real if and only if \mbox{$h(r)^{2}a'(r) \equiv 0$.}  Since $h$ has only isolated zeros, $h(r)^{2}a'(r) \equiv 0$ if and only if $a'(r) \equiv 0$ or equivalently $a(r)$ is constant.
\end{proof}

Note that if $a \neq 0$, then the resulting solution, $f(t,r) = \e^{i\om t}\e^{ia}h(r)$ is equivalent to the usual ansatz of a standing wave $f(t,r) = \e^{i\om t} F(r)$ with $F$ real valued if the prior solution starts at \mbox{$t=-a/\om$} instead of $t=0$.  That is, $a\neq 0$ is simply a $t$-translation of the usual standing wave solutions in \cite{Bizon00}.

\section{Families of Static States}

The more important comment in this paper concerns a few interesting relationships of the parameters defining static states.  Substituting equation (\ref{sfstaticstate}) with $F$ real-valued into the system (\ref{EKG}) yields the first order system of ODEs
\begin{subequations} \label{NCpdec}
  \begin{align}
    \label{NCpde1c} M' &= 4\pi r^{2}\mu_{0}\brkt{\pnth{1 + \frac{\om^{2}}{\Up^{2}}\e^{-2V}}\abs{F}^{2} + \pnth{1 - \frac{2M}{r}}\frac{\abs{H}^{2}}{\Up^{2}}}, \\
    \label{NCpde2c} V' &= \pnth{1 - \frac{2M}{r}}^{-1}\brce{\frac{M}{r^{2}} - 4\pi r\mu_{0}\brkt{\pnth{1 - \frac{\om^{2}}{\Up^{2}}\e^{-2V}}\abs{F}^{2} - \pnth{1 - \frac{2M}{r}}\frac{\abs{H}^{2}}{\Up^{2}}}}, \\
    \label{NCpde3c} F' &= H, \\
    \label{NCpde4c} H' &= \pnth{1 - \frac{2M}{r}}^{-1}\brkt{\pnth{\Up^{2} - \frac{\om^{2}}{\e^{2V}}} F + 2H\pnth{\frac{M}{r^{2}} + 4\pi r\mu_{0}\abs{F}^{2} - \frac{1}{r}}},
  \end{align}
\end{subequations}
where $H$ is defined by equation (\ref{NCpde3c}).  To solve these equations numerically, we need boundary conditions both at $r=0$ and $r=r_{max}$, the largest $r$-value included in the computation.

For regularity at $r=0$, $H(0)=0$ and $M(0)=0$.  Note that if $F(0)=F_{0} \neq 0$ (if $F(0)=0$, then $F \equiv 0$ \cite{ParryPhD}), then we can factor $F_{0}$ out of $f$ in equations (\ref{EKG-1}) and (\ref{EKG-2}).  In (\ref{EKG-2}), $F_{0}$ cancels and in (\ref{EKG-1}), $(F_{0})^{2}$ can be absorbed into the value of $\mu_{0}$.  This yields another solution to the system (\ref{EKG}) qualitatively equivalent to the first, but with $F(0)=1$.  Thus, without loss of generality, set $F(0)=1$.

Next we consider the behavior of the functions at the outer boundary.  Since $N$ is asymptotically Schwarzschild, there exist constants, $m\geq 0$, called the total mass of the system, and $\ka >0$, and a Schwarzschild metric $g_{S}$ given by
\begin{equation}\label{SW-metric}
  g_{S} = -\ka^{2}\pnth{1 - \frac{2m}{r}}\, dt^{2} + \pnth{1 - \frac{2m}{r}}^{-1}\, dr^{2} + r^{2}\, d\si^{2},
\end{equation}
such that $g$ approaches $g_{S}$ as $r \to \infty$.  This yields the following asymptotic boundary conditions.
\begin{align}
  \label{AB1a} \glap{g_{S}}f &\to \Upsilon^2 f \quad \text{and} \quad f \to 0 \quad \text{as } r \to \infty, \\
  \label{AB2a} \e^{2V} &\to \ka^{2}\pnth{1 - \frac{2M}{r}} \quad \text{as } r \to \infty.
\end{align}
The first boundary condition (\ref{AB1a}) implies by equation (\ref{NCpde1a}) that $M_{r} \to 0$ as $r \to \infty$ and hence $M$ approaches a constant value as $r \to \infty$. Given equations (\ref{metric}), (\ref{SW-metric}), and the second boundary condition (\ref{AB2a}), this constant will be the parameter $m$ in (\ref{SW-metric}).  To deal with these equations numerically, we will impose these conditions at $r=r_{max}$.

In this case, equation (\ref{AB2a}) becomes
\begin{align}
  \notag \e^{2V(r_{max})} &\approx \ka^{2}\pnth{1 - \frac{2M(r_{max})}{r_{max}}} \\
  \label{AB2b} 0 &\approx V(r_{max}) - \frac{1}{2}\ln\pnth{1 - \frac{2M(r_{max})}{r_{max}}} - \ln \ka.
\end{align}

For (\ref{AB1a}) at $r = r_{max}$, we require $f$ to approximately solve the Klein-Gordon equation (\ref{EKG-2}) in the Schwarzschild metric (\ref{SW-metric}).  Computing the Laplacian in the Schwarzschild metric, this equation becomes
\begin{equation}\label{lapS1}
  \pnth{1 - \frac{2m}{r}}^{2}F'' + \pnth{1 - \frac{2m}{r} + \pnth{1 - \frac{2m}{r}}^{2}}\frac{F'}{r} - \pnth{\Up^{2}\pnth{1 - \frac{2m}{r}} - \frac{\om^{2}}{\ka^{2}}}F = 0.
\end{equation}
For large $r$, this simplifies to
\begin{equation}\label{lapS2}
  F'' + \frac{2F'}{r} - \pnth{\Up^{2} - \frac{\om^{2}}{\ka^{2}}}F = 0.
\end{equation}
This ODE is routinely solved and has the general solution
\begin{equation}\label{Ssol}
  F = \frac{C_{1}}{r}\e^{r\sqrt{\Up^{2} - \frac{\om^{2}}{\ka^{2}}}} + \frac{C_{2}}{r}\e^{-r\sqrt{\Up^{2}-\frac{\om^{2}}{\ka^{2}}}}
\end{equation}
for some constants $C_{1},C_{2} \in \RR$.  Now equation \ref{AB1a} also requires that $F \to 0$ as $r \to \infty$ so that $f \to 0$ as well.  Thus $C_{1} = 0$ and we relabel $C_{2}$ as simply $C$.  That is, at $r = r_{max}$, we require
\begin{equation}\label{pre-AB1}
  F = \frac{C}{r}\e^{-r\sqrt{\Up^{2} - \frac{\om^{2}}{\ka^{2}}}}.
\end{equation}
We have no way of directly determining the correct value of $C$ in equation (\ref{pre-AB1}) associated with a given static solution.  However, if we differentiate equations (\ref{pre-AB1}) with respect to $r$, we obtain
\begin{align}
  \notag F' &= -\frac{C}{r}\e^{-r\sqrt{\Up^{2} - \frac{\om^{2}}{\ka^{2}}}}\sqrt{\Up^{2} - \frac{\om^{2}}{\ka^{2}}} - \frac{C}{r^{2}}\e^{-r\sqrt{\Up^{2} - \frac{\om^{2}}{\ka^{2}}}} \\
  &= -\pnth{\sqrt{\Up^{2} - \frac{\om^{2}}{\ka^{2}}} + \frac{1}{r}}F,
\end{align}
which does not depend on $C$.  Then the condition that at $r=r_{max}$, $f$ approximately satisfies the Klein-Gordon equation with the Schwarzschild background metric reduces to requiring that
\begin{equation}\label{AB1b}
  F'(r_{max}) + \pnth{\sqrt{\Up^{2} - \frac{\om^{2}}{\ka^{2}}} + \frac{1}{r_{max}}}F(r_{max}) \approx 0.
\end{equation}
Essentially, this condition imposes that $F$ is decaying appropriately to $0$.  It also puts a restriction on the possible values of $\om$.  Since the left hand side of the above equation must be real, we have that $\Up^{2} - \frac{\om^{2}}{\ka^{2}} \geq 0$, or equivalently,
\begin{equation}
  \abs{\frac{\om}{\ka}} = \frac{\om}{\ka} \leq \Up.
\end{equation}
That is, $\om/\ka \in [0,\Up]$.  For simplicity in our calculations, we set $\ka = 1$ so that the metric is asymptotic to the standard Schwarzschild solution.

With both of these boundary conditions in mind, we see that the parameters $\Up$, $\mu_{0}$, $\om$, and $V(0)=V_{0}$ define solutions to the system (\ref{NCpdec}).  Then we solve this system as follows.  For some choice of $\Up$ and $\mu_{0}$, we require at $r = 0$ that
\begin{align}
  \label{cenval} F(0) &= 1, & H(0) &= 0, & M(0) &= 0, & V(0) &= V_{0},
\end{align}
and choose $\om$ and $V_{0}$ via solving a shooting problem to satisfy
\begin{align}
  \label{AB1c} F'(r_{max}) + \pnth{\sqrt{\Up^{2} - \om^{2}} + \frac{1}{r_{max}}}F(r_{max}) &\approx 0, \\
  \label{AB2c} V(r_{max}) - \frac{1}{2}\ln\pnth{1 - \frac{2M(r_{max})}{r_{max}}} &\approx 0.
\end{align}
See \cite{ParryPhD} for a more detailed computation of these ODEs and boundary conditions.

\begin{figure}

  \begin{center}
    \includegraphics[height = 1.5 in, width = 2 in]{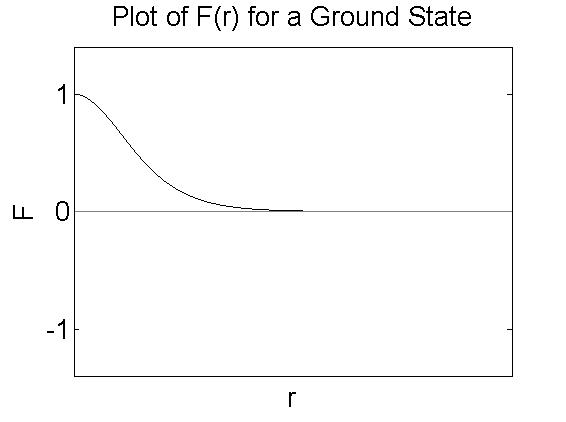}
    \includegraphics[height = 1.5 in, width = 2 in]{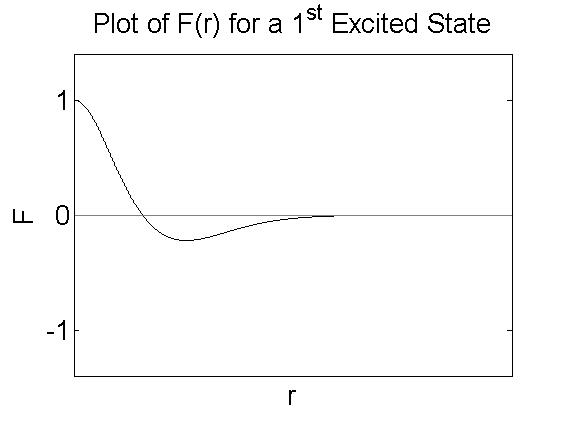}
    \includegraphics[height = 1.5 in, width = 2 in]{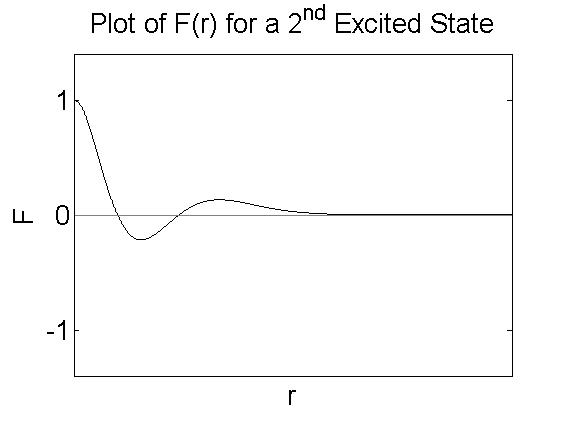}
  \end{center}

  \caption{Plots of static state scalar fields (specifically the function $F(r)$ in (\ref{sfstaticstate})) in the ground state and first and second excited states.  Note the number of nodes (zeros) of each function.}

  \label{state_ex}

\end{figure}

\begin{figure}

    \begin{center}
        \includegraphics[height = 1.5 in, width = 2 in]{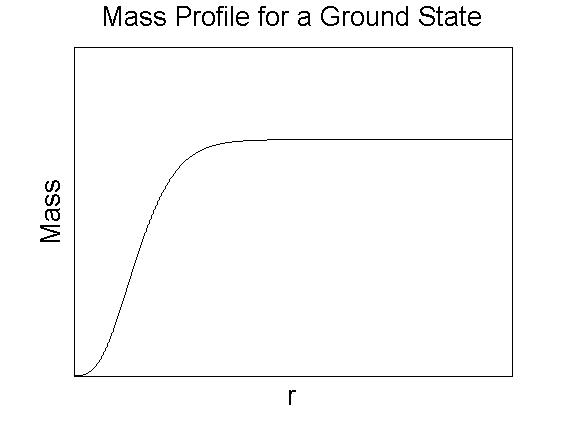}
        \includegraphics[height = 1.5 in, width = 2 in]{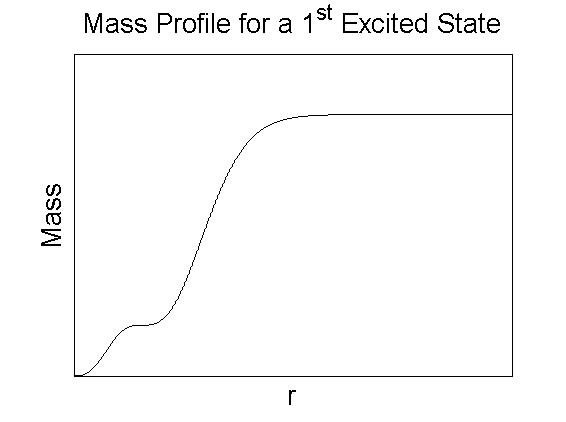}
        \includegraphics[height = 1.5 in, width = 2 in]{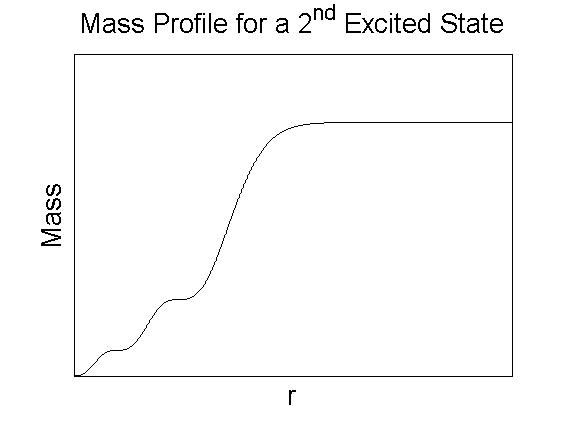}
    \end{center}

    \caption{Mass profiles for a static ground state and first and second excited states.}

    \label{state_m}

\end{figure}

Since equations (\ref{AB1c}) and (\ref{AB2c}) are necessary for the system to make 
physical sense, the choice of $\Up$ and $\mu_{0}$ completely define the set of static states.  It is discussed in \cite{Bizon00}, albeit with different notation, that a choice of $\Up$ and $\mu_{0}$ corresponds to a countable set of solutions where each member can be distinquished by the number of nodes $n$ it has.  An example of a ground state (n=0), first excited state (n=1), and second excited state (n=2) is depicted in figure \ref{state_ex} with corresponding plots of the mass function $M$ in figure \ref{state_m}.

\begin{figure}

    \begin{center}
        \includegraphics[height = 1.5 in, width = 2 in]{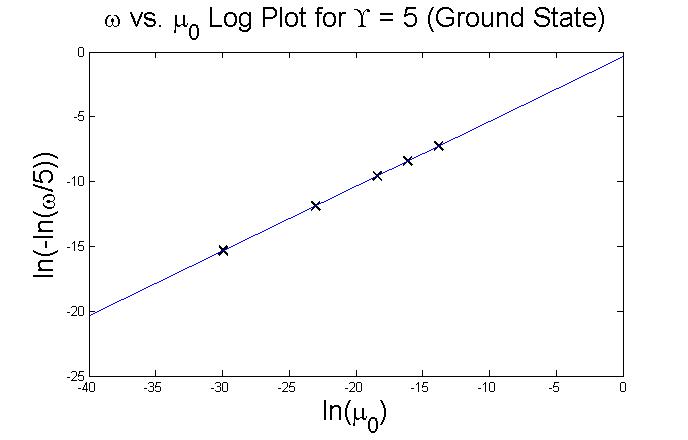}
        \includegraphics[height = 1.5 in, width = 2 in]{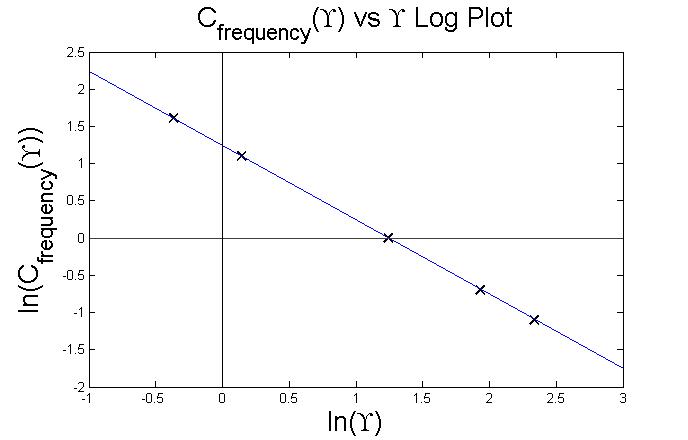}
    \end{center}

    \caption{Left: Log plot of the parameters $\om$ and $\mu_{0}$ for a ground state and constant value of $\Up = 5$.  The slope of this plot is almost exactly $1/2$.  We get the same slope for other values of $\Up$, thus $\om(\Up,\mu_{0}) = \Up \e^{C_{frequency}(\Up)\sqrt{\mu_{0}}}$.  Right: Log plot of the parameters $C_{frequency}(\Up)$ and $\Up$ for a ground state.  The slope of this plot is almost exactly $-1$.  Thus $C_{frequency}(\Up) = C_{frequency}/\Up$, where $C_{frequency}$ is a constant.  Similar plots exist for any $n^{\text{th}}$ excited state.}

    \label{logplot}

\end{figure}

One might ask if there is some explicit relationship between the choice of $\Up$ and $\mu_{0}$ and the choice of $\om$ and $V_{0}$ and it turns out that there is.   In fact, we have also found expressions for the total mass $m$ of the system (the constant to which the mass function $M$ is asymptotic) and the half-mass radius $r_{h}$, that is, the radius $r_{h}$ for which $M(r_{h}) = m/2$. These expressions were found by numerically computing static states for several different values of $\Up$ and $\mu_{0}$, all of which yielded states in the low field limit.  We analyzed the resulting values and found that certain log plots between the values were linear.  We have collected in Figure \ref{logplot} some of these plots for the ground state that led to this conclusion.

These log plots yielded the following expressions, which we emphasize are only expected to hold only in the low field limit.  Let $\om^{n}$, $V_{0}^{n}$, $m^{n}$, $r_{h}^{n}$ be respectively the values of $\om$, $V(0)$, $m$, and $r_{h}$ for an $n^{\th}$-excited state corresponding to a choice of $\Up$ and $\mu_{0}$.  Then we have that
\begin{subequations}\label{Scorr}
\begin{align}
  \label{Scorr1} \om^{n}(\Up,\mu_{0}) &\approx \Up \exp\pnth{C_{frequency}^{n}\frac{\sqrt{\mu_{0}}}{\Up}}, \\
  \label{Scorr2} V_{0}^{n}(\Up,\mu_{0}) &\approx C_{potential}^{n}\frac{\sqrt{\mu_{0}}}{\Up}, \\
  \label{Scorr3} m^{n}(\Up,\mu_{0}) &\approx C_{mass}^{n}\Up^{-3/2}\mu_{0}^{1/4}, \\
  \label{Scorr4} r_{h}^{n}(\Up,\mu_{0}) &\approx C_{radius}^{n}\Up^{-1/2}\mu_{0}^{-1/4},
\end{align}
\end{subequations}
for some constants $C_{frequency}^{n}$, $C_{potential}^{n}$, $C_{mass}^{n}$ and $C_{radius}^{n}$ which depend only on $n$.  We have computed these constants for the ground through fifth excited states as well as for the tenth and twentieth excited states and have collected these values in Table \ref{ScorrVal}.

\begin{table}

  \begin{center}

  \begin{tabular}{c|cccc}
    $n$ & $C_{frequency}^{n}$ & $C_{potential}^{n}$ & $C_{mass}^{n}$ & $C_{radius}^{n}$ \\
    \hline & & & & \\
    $0$ & $-3.4638 \pm 0.010$ & $-6.7278 \pm 0.003$ & $4.567 \pm 0.05$ & $0.8462 \pm 0.004$ \\
    $1$ & $-3.2422 \pm 0.012$ & $-7.5411 \pm 0.007$ & $10.22 \pm 0.10$ & $2.2894 \pm 0.009$ \\
    $2$ & $-3.1566 \pm 0.014$ & $-7.9315 \pm 0.009$ & $15.81 \pm 0.16$ & $3.8253 \pm 0.014$ \\
    $3$ & $-3.1062 \pm 0.015$ & $-8.1823 \pm 0.010$ & $21.37 \pm 0.22$ & $5.3994 \pm 0.018$ \\
    $4$ & $-3.0714 \pm 0.015$ & $-8.3653 \pm 0.010$ & $26.91 \pm 0.27$ & $6.9860 \pm 0.022$ \\
    $5$ & $-3.0452 \pm 0.016$ & $-8.5086 \pm 0.011$ & $32.42 \pm 0.33$ & $8.5606 \pm 0.026$ \\
    $10$ & $-3.0076 \pm 0.052$ & $-9.0018 \pm 0.037$ & $60.32 \pm 1.18$ & $15.1357 \pm 0.039$ \\
    $20$ & $-2.9949 \pm 0.077$ & $-9.5061 \pm 0.074$ & $116.62 \pm 2.57$ & $29.6822 \pm 0.107$
  \end{tabular}

  \caption{Values of the constants in the system (\ref{Scorr}) for the ground through fifth excited states as well as the tenth and twentieth excited states.  We have given them error ranges which encompass the interval we observed in our experiments.  However, it is possible that values outside our ranges here could be observed, though we don't expect them to be so by much if the discretization of $r$ used in solving the ODEs is sufficiently fine.  Note also that our values have less precision as we increase $n$.  This is because as $n$ increases, it becomes more difficult to compute the states with as much precision.}

  \label{ScorrVal}

  \end{center}

\end{table}

The above approximations only hold in the low field limit and are consistent with the Einstein-Klein-Gordon equations in the low field limit.  Specifically, the following approximations represent the low field limit in our current setup,
\begin{subequations}\label{LFLapprox}
  \begin{align}
    \label{LFLapprox1} \pnth{1 - \frac{2M}{r}} &\approx 1, & \e^{V} &\approx 1, \\
    \label{LFLapprox2} \frac{\om}{\Up} &\approx 1, & \frac{H}{\Up} &\approx 0.
  \end{align}
\end{subequations}
This is due to the fact that being in the low field limit is essentially the statement that our system is a perturbation of the Minkowski spacetime.  The line (\ref{LFLapprox1}) is the statement that the metric (\ref{metric}) is a perturbation of the Minkowski metric, which includes claiming that the mass is small.    Note also the approximation $\e^{V} \approx 1$ makes valid the Taylor polynomial approximation $\e^{cV} \approx 1 + cV$ if $c$ is not too large.  To explain the next approximation, note that, in Minkowski space, the equation $\glap{g}f = \Up^{2} f$ is the equation
\begin{equation}
  -f_{tt} + f_{rr} + \frac{2f_{r}}{r} = \Up^{2} f.
\end{equation}
Our system should approximately satsify this equation if it is a perturbation of Minkowski space.  For our system to be in the low field limit, it is natural to require that $f$, which controls the stress energy tensor, does not change too quickly in $r$ at least on the scale of $\Up$.  This translates into $f_{r} \approx 0$ and $f_{rr} \approx 0$ on the scale of $\Up$.  This makes the above equation
\begin{equation}\label{MinkKGapprox}
  -f_{tt} \approx \Up^{2}f.
\end{equation}
Given our ansatz (\ref{sfstaticstate}), equation (\ref{MinkKGapprox}) is satisfied when $\om \approx \Up$, which is equivalent to the first approximation in line (\ref{LFLapprox2}) above.  Lastly, since $F' = H$, the final approximation in line (\ref{LFLapprox2}) is the statement that $F$ does not change much on the scale of $\Up$, which given equation (\ref{sfstaticstate}) is the same as our previous requirement that $f$ not change much on the scale of $\Up$.

Applying these approximations to equation (\ref{NCpde1c}) yields the following
\begin{align}
  \notag M' &= 4\pi r^{2}\mu_{0}\brkt{\pnth{1 + \frac{\om^{2}}{\Up^{2}}\e^{-2V}}\abs{F}^{2} + \pnth{1 - \frac{2M}{r}}\frac{\abs{H}^{2}}{\Up^{2}}} \\
  \notag &\approx 4\pi r^{2}\mu_{0}\brkt{\pnth{1 + \e^{-2V}}\abs{F}^{2}} \\
  \notag &\approx 4\pi r^{2}\mu_{0}\brkt{2\abs{F}^{2}} \\
  \label{PS1a} &= 8\pi r^{2}\mu_{0}\abs{F}^{2}.
\end{align}
Since $M' = 4\pi r^{2} \mu$, this yields that the low field limit of the energy density is $2\mu_{0}\abs{F}^{2}$.  Applying the approximations to equation (\ref{NCpde2c}) yields
\begin{align}
  \notag V' &= \pnth{1 - \frac{2M}{r}}^{-1}\brce{\frac{M}{r^{2}} - 4\pi r\mu_{0}\brkt{\pnth{1 - \frac{\om^{2}}{\Up^{2}}\e^{-2V}}\abs{F}^{2} - \pnth{1 - \frac{2M}{r}}\frac{\abs{H}^{2}}{\Up^{2}}}} \\
  \notag &\approx \frac{M}{r^{2}} - 4\pi r\mu_{0}\brkt{\pnth{1 - \e^{-2V}}\abs{F}^{2}} \\
  \label{PS1b} &\approx \frac{M}{r^{2}}.
\end{align}
This approximation coupled with the previous approximation imply that the Poisson equation
\begin{equation}\label{PS1}
  \lap_{\RR^{3}}V = 4\pi\pnth{2\mu_{0}\abs{F}^{2}}
\end{equation}
is approximately true.

The last two equations (\ref{NCpde3c}) and (\ref{NCpde4c}) can be rewritten as follows.
\begin{align}
  \notag F'' &= \pnth{1 - \frac{2M}{r}}^{-1}\brkt{\pnth{\Up^{2} - \frac{\om^{2}}{\e^{2V}}} F + 2F'\pnth{\frac{M}{r^{2}} + 4\pi r\mu_{0}\abs{F}^{2} - \frac{1}{r}}} \\
  \notag &\approx \pnth{\Up^{2} - \frac{\om^{2}}{\e^{2V}}} F + 2F'\pnth{\frac{M}{r^{2}} + 4\pi r\mu_{0}\abs{F}^{2} - \frac{1}{r}} \displaybreak[0] \\
 \notag  F'' + \frac{2F'}{r} &\approx \pnth{\Up^{2} - \om^{2}(1 - 2V)}F + 2F'\pnth{\frac{M}{r^{2}} + 4\pi r\mu_{0}\abs{F}^{2}} \displaybreak[0] \\
  \notag \lap_{\RR^{3}}F & \approx \pnth{\Up^{2} - \om^{2}}F + 2\om^{2}VF + 2F'\pnth{\frac{M}{r^{2}} + 4\pi r\mu_{0}\abs{F}^{2}} \displaybreak[0] \\
  \notag & \approx \pnth{\Up + \om}\pnth{\Up - \om}F + 2\om^{2}VF + 2F'\pnth{\frac{M}{r^{2}} + 4\pi r\mu_{0}\abs{F}^{2}} \displaybreak[0] \\
  \notag & \approx 2\Up\pnth{\Up - \om}F + 2\om^{2}VF + 2F'\pnth{\frac{M}{r^{2}} + 4\pi r\mu_{0}\abs{F}^{2}} \displaybreak[0] \\
  \notag \frac{1}{2\Up}\lap_{\RR^{3}}F & \approx \pnth{\Up - \om}F + \frac{\om^{2}}{\Up}VF + \frac{F'}{\Up}\pnth{\frac{M}{r^{2}} + 4\pi r\mu_{0}\abs{F}^{2}} \displaybreak[0] \\
  \notag & \approx \pnth{\Up - \om}F + \frac{\om^{2}}{\Up}VF + \frac{F'}{\Up}\pnth{\frac{M}{r^{2}} + 4\pi r\mu_{0}\abs{F}^{2}} \displaybreak[0] \\
  \notag & \approx \pnth{\Up - \om}F + \om VF \\
  \label{PS2} \frac{1}{2\Up}\lap_{\RR^{3}}F & \approx \pnth{\Up - \om}F + \Up VF
\end{align}
where the last line is because $\om \approx \Up$.

Under these assumptions, the system (\ref{NCpdec}) reduces to 
\begin{subequations}\label{LFLNCpdec}
  \begin{align}
    \label{LFLNCpde1c} M' &\approx 8\pi r^{2}\mu_{0}\abs{F}^{2} \\
    \label{LFLNCpde2c} V' &\approx \frac{M}{r^{2}} \\
    \label{LFLNCpde4c} \frac{1}{2\Up}\lap_{\RR^{3}}F &\approx (\Up - \om)F + \Up V F.
  \end{align}
\end{subequations}
These are essentially the Poisson-\Schrodinger equations, which are a well known low field limit version of the Einstein-Klein-Gordon equations.  Equations (\ref{LFLNCpde1c}) and (\ref{LFLNCpde2c}) imply the Poisson equation $\lap_{\RR^{3}}V = 4\pi \mu$ and equation (\ref{LFLNCpde4c}) is the \Schrodinger equation in this setup.  The set of approximations (\ref{Scorr}) are consistent with the system (\ref{LFLNCpdec}) in that solutions scale appropriately.  For a given state and constant $\Up$, the approximations (\ref{Scorr}) describe a one parameter ($\mu_{0}$) family of solutions to the Einstein-Klein-Gordon equations in the low field limit.  If we replace $\mu_{0}$ by $\al^{4}\mu_{0}$ for some $\al >0$, then equations (\ref{Scorr}) suggest that the values $\Up - \om$, $V$, $M$, and $r$ scale as follows
\begin{subequations}\label{scalings}
  \begin{align}
    \mu_{0} &= \al^{4}\hat{\mu}_{0}, & V &= \al^{2}\hat{V}, \\
    M &\to \al \hat{M}, & r &\to \frac{1}{\al}\hat{r}, \\    
    \Up - \om &\to \al^{2}(\Up - \hat{\om}). \\
  \end{align}
\end{subequations}
The last scaling is due to the approximation $\e^{x} \approx 1 + x$ which holds for equation (\ref{Scorr1}) as long as $\mu_{0}$ is small.  Note that since $r = \frac{1}{\al}\hat{r}$, every derivative multiplies the scale factor by $\al$.  If we do not scale $F$ and make these substitutions in the system (\ref{LFLNCpdec}), we obtain
\begin{subequations}\label{LFLscale}
  \begin{align}
    \notag \al^{2}\hat{M}' &\approx 8\pi \frac{1}{\al^{2}}\hat{r}^{2}\al^{4}\hat{\mu}_{0}\abs{F}^{2} \\
    \label{LFLscale1} \hat{M}' &\approx 8\pi \hat{r}^{2}\hat{\mu}_{0}\abs{F}^{2} \\
    \notag \al^{3}\hat{V}' &\approx \frac{\al\hat{M}}{\hat{r}^{2}/\al^{2}} = \al^{3}\frac{\hat{M}}{\hat{r}^{2}} \\
    \label{LFLscale2} \hat{V}' &\approx \frac{\hat{M}}{\hat{r}^{2}} \\
    \notag \frac{1}{2\Up}\al^{2}\lap_{\RR^{3}}F &\approx \al^{2}(\Up - \hat{\om})F + \Up \al^{2}\hat{V} F \\
    \label{LFLscale3} \frac{1}{2\Up}\lap_{\RR^{3}}F &\approx (\Up - \hat{\om})F + \Up \hat{V} F.
  \end{align}
\end{subequations}
Thus the Einstein-Klein-Gordon equations in the low field limit are approximately invariant under the scalings implied by the approximations (\ref{Scorr}).

\begin{figure}

    \begin{center}
      \includegraphics[width=3 in]{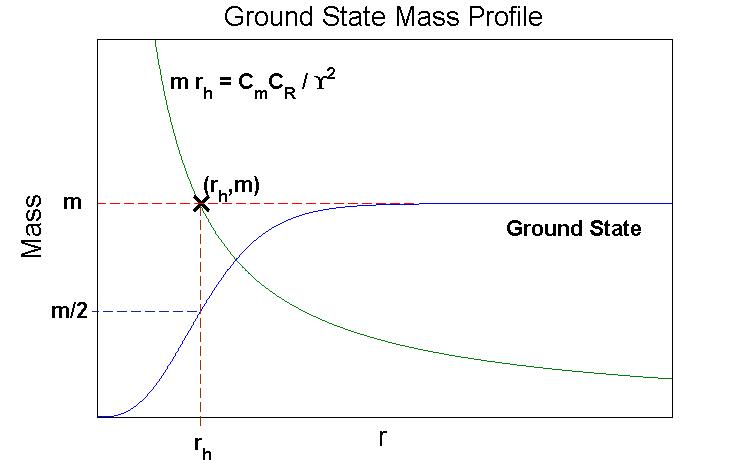}
      \includegraphics[width=3 in]{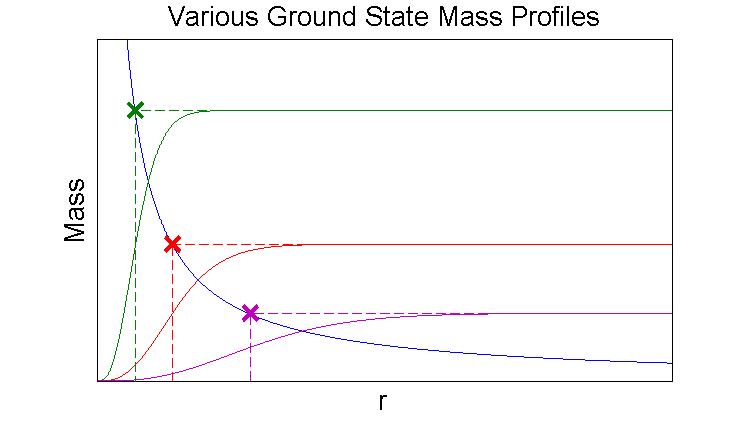}
    \end{center}

    \caption{Left: Plot of the mass profile of a ground state with its corresponding hyperbola of constant $\Up$ overlayed.  Any ground state mass profile that keeps the presented relationship with this hyperbola corresponds to the same value of $\Up$.  Right: Examples of different ground state mass profiles corresponding to the same value of $\Up$.  The corresponding hyperbola of constant $\Up$ is overlayed.  Notice that all three mass profiles have the same relationship with the hyperbola.}

    \label{mass-hyperbola-figure}

\end{figure}
    
A remarkable property of the system (\ref{Scorr}) is the relationship it produces between $\Up$, $m$, and $r_{h}$.  We observe from equations (\ref{Scorr3}) and (\ref{Scorr4}) that the product $mr_{h}$ does not depend on the parameter $\mu_{0}$.  Specifically,
\begin{equation}\label{mass-hyperbola}
    m r_{h} = \frac{C_{mass}C_{radius}}{\Up^{2}},
\end{equation}
where we have suppressed the notation of $n$.  If $\Up$ is constant, then, because both $C_{mass}$ and $C_{radius}$ are positive for all $n$ (see Table \ref{ScorrVal}), the right hand side of this equation is some positive constant, $k$, and we have that
\begin{equation}
  m r_{h} = k,
\end{equation}
which defines a hyperbola.  Thus, for a given $n^{\text{th}}$ excited state, all of the possible mass profiles for a constant value of $\Up$ lie along a hyperbola.  We illustrate this phenomenon in Figure \ref{mass-hyperbola-figure}.  This has important implications in being able to determine values of $\Up$ that can produce physically plausible mass profiles.

\section{Outlook}

The remarks made in this paper are useful in understanding the implications of wave dark matter in the case where the spacetime is static and spherically symmetric.  In particular, the fact that the mass profiles of static states lie on hyperbolas of constant $\Up$ can be used to test various values of $\Up$ against actual data and contribute to the determining of whether wave dark matter is a viable dark matter candidate.

\section{Acknowledgements}

The author would like to thank Hubert Bray and Andrew Goetz for many helpful discussions and for invaluable suggestions about the content of this paper.  He also gratefully acknowledges the support of the National Science Foundation grant \#~DMS-1007063.

\bibliographystyle{phaip}
\bibliography{./References}

\begin{bibdiv}
\begin{biblist}

\bib{Seidel98}{article}{
      author={Balakrishna, Jayashree},
      author={Seidel, Edward},
      author={Suen, Wai-Mo},
       title={{Dynamical} {Evolution} of {Boson} {Stars}. {I}{I}. {Excited}
  {States} and {Self-interacting} {Fields}},
        date={1998-09},
     journal={Phys. Rev. D},
      volume={58},
       pages={104004},
      eprint={http://link.aps.org/doi/10.1103/PhysRevD.58.104004},
}

\bib{MSBS}{article}{
      author={Bernal, A.},
      author={Barranco, J.},
      author={Alic, D.},
      author={Palenzuela, C.},
       title={Multistate {Boson} {Stars}},
        date={2010-02},
     journal={Phys. Rev. D},
      volume={81},
      number={4},
       pages={044031},
      eprint={http://arxiv.org/abs/0908.2435},
}

\bib{Bernal08}{article}{
      author={Bernal, A.},
      author={Matos, T.},
      author={{N\'{u}\~{n}ez}, D.},
       title={{Flat Central Density Profiles from Scalar Field Dark Matter
  Halos}},
        date={2008},
     journal={Revista Mexicana de {Astronom\'{i}a} y {Astrof\'{i}sica}},
      volume={44},
       pages={149\ndash 160},
      eprint={http://arxiv.org/abs/astro-ph/0303455v3},
}

\bib{Bizon00}{article}{
      author={Bizo\'{n}, Piotr},
      author={Wasserman, Arthur},
       title={On existence of mini-boson stars},
    language={English},
        date={2000},
        ISSN={0010-3616},
     journal={Communications in Mathematical Physics},
      volume={215},
      number={2},
       pages={357\ndash 373},
         url={http://dx.doi.org/10.1007/s002200000307},
}

\bib{Bray12}{article}{
      author={Bray, H.},
       title={On {Wave} {Dark} {Matter}, {Shells} in {Elliptical} {Galaxies},
  and the {Axioms} of {General} {Relativity}},
        date={2012-12},
     journal={[arXiv:1212.5745 [physics.gen-ph]]},
      eprint={http://arxiv.org/abs/1212.5745},
}

\bib{Bray10}{article}{
      author={Bray, H.},
       title={On {Dark} {Matter}, {Spiral} {Galaxies}, and the {Axioms} of
  {General} {Relativity}},
        date={2013},
     journal={AMS Contemporary Mathematics},
      volume={599},
       pages={1\ndash 64},
      eprint={http://arxiv.org/abs/1004.4016},
}

\bib{Feinblum68}{article}{
      author={Feinblum, David~A.},
      author={McKinley, William~A.},
       title={Stable states of a scalar particle in its own gravational field},
        date={1968Apr},
     journal={Phys. Rev.},
      volume={168},
       pages={1445\ndash 1450},
         url={http://link.aps.org/doi/10.1103/PhysRev.168.1445},
}

\bib{Guzman00}{article}{
      author={{Guzm\'{a}n}, F.~S.},
      author={Matos, T.},
       title={Scalar fields as dark matter in spiral galaxies},
        date={2000},
     journal={Classical and Quantum Gravity},
      volume={17},
      number={1},
       pages={L9},
      eprint={http://stacks.iop.org/0264-9381/17/i=1/a=102},
}

\bib{Guzman01}{article}{
      author={{Guzm\'{a}n}, F.~S.},
      author={Matos, T.},
      author={Villegas-Brena, H.},
       title={Scalar {Dark} {Matter} in {Spiral} {Galaxies}},
        date={2001},
     journal={Rev. Mex. Astron. Astrofis.},
      volume={37},
       pages={63\ndash 72},
      eprint={http://arxiv.org/abs/astro-ph/9811143},
}

\bib{Jetzer92B}{article}{
      author={Jetzer, Phillippe},
       title={Boson stars},
        date={1992},
        ISSN={0370-1573},
     journal={Physics Reports},
      volume={220},
      number={4},
       pages={163\ndash 227},
  url={http://www.sciencedirect.com/science/article/pii/037015739290123H},
}

\bib{Ji94}{article}{
      author={Ji, S.~U.},
      author={Sin, S.~J.},
       title={Late-{Time} {Phase} {Transition} and the {Galactic} {Halo} as a
  {Bose} {Liquid}. {II}. {The} {Effect} of {Visible} {Matter}},
        date={1994-09},
     journal={Phys. Rev. D},
      volume={50},
       pages={3655\ndash 3659},
      eprint={http://link.aps.org/doi/10.1103/PhysRevD.50.3655},
}

\bib{Kaup68}{article}{
      author={Kaup, David~J.},
       title={Klein-gordon geon},
        date={1968Aug},
     journal={Phys. Rev.},
      volume={172},
       pages={1331\ndash 1342},
         url={http://link.aps.org/doi/10.1103/PhysRev.172.1331},
}

\bib{Kusmart91}{article}{
      author={Kusmartsev, Fjodor~V.},
      author={Mielke, Eckehard~W.},
      author={Schunck, Franz~E.},
       title={Gravitational stability of boson stars},
        date={1991Jun},
     journal={Phys. Rev. D},
      volume={43},
       pages={3895\ndash 3901},
         url={http://link.aps.org/doi/10.1103/PhysRevD.43.3895},
}

\bib{Lee09}{article}{
      author={Lee, J.-W.},
       title={Is {Dark} {Matter} a {BEC} or {Scalar} {Field}?},
        date={2009},
     journal={Journal of the Korean Physical Society},
      volume={54},
      number={6},
       pages={2622},
      eprint={http://arxiv.org/abs/0801.1442},
}

\bib{Lee92}{article}{
      author={Lee, J.-W.},
      author={Koh, I.-G.},
       title={Galactic {Halo} as a {Soliton} {Star}},
        date={1992},
     journal={Abstracts, Bulletin of the Korean Physical Society},
      volume={10},
      number={2},
}

\bib{Lee96}{article}{
      author={Lee, J.-W.},
      author={Koh, I.-G.},
       title={Galactic {Halos} as {Boson} {Stars}},
        date={1996-02},
     journal={Phys. Rev. D},
      volume={53},
       pages={2236\ndash 2239},
      eprint={http://link.aps.org/doi/10.1103/PhysRevD.53.2236},
}

\bib{Matos09}{article}{
      author={Matos, Tonatiuh},
      author={{V\'{a}zquez-Gonz\'{a}lez}, Alberto},
      author={{Maga\~{n}a}, Juan},
       title={$\varphi^{2}$ as dark matter},
        date={2009},
        ISSN={1365-2966},
     journal={Monthly Notices of the Royal Astronomical Society},
      volume={393},
      number={4},
       pages={1359\ndash 1369},
      eprint={http://dx.doi.org/10.1111/j.1365-2966.2008.13957.x},
}

\bib{ParryPhD}{thesis}{
      author={Parry, Alan~R.},
       title={Wave {Dark} {Matter} and {Dwarf} {Spheroidal} {Galaxies}},
        type={Ph.D. Thesis},
        date={2013},
}

\bib{Parry12-1}{article}{
      author={Parry, Alan~R.},
       title={A {Survey} of {Spherically} {Symmetric} {Spacetimes}},
    language={English},
        date={2014},
        ISSN={1664-2368},
     journal={Analysis and Mathematical Physics},
      volume={4},
      number={4},
       pages={333\ndash 375},
      eprint={http://dx.doi.org/10.1007/s13324-014-0085-x},
         url={http://dx.doi.org/10.1007/s13324-014-0085-x},
}

\bib{Pugliese13}{article}{
      author={Pugliese, Daniela},
      author={Quevedo, Hernando},
      author={Rueda~H., Jorge~A.},
      author={Ruffini, Remo},
       title={Charged boson stars},
        date={2013Jul},
     journal={Phys. Rev. D},
      volume={88},
       pages={024053},
         url={http://link.aps.org/doi/10.1103/PhysRevD.88.024053},
}

\bib{Mielke03}{article}{
      author={Schunck, F.},
      author={Mielke, E.},
       title={General relativistic boson stars},
        date={2003},
     journal={Classical and Quantum Gravity},
      volume={20},
       pages={R301\ndash R356},
      eprint={http://stacks.iop.org/0264-9381/20/i=20/a=201},
}

\bib{Schunck98}{article}{
      author={Schunck, Franz~E.},
      author={Mielke, Eckehard~W.},
       title={Rotating boson star as an effective mass torus in general
  relativity},
        date={1998},
        ISSN={0375-9601},
     journal={Physics Letters A},
      volume={249},
      number={5–6},
       pages={389\ndash 394},
  url={http://www.sciencedirect.com/science/article/pii/S0375960198007786},
}

\bib{Seidel90}{article}{
      author={Seidel, Edward},
      author={Suen, Wai-Mo},
       title={{Dynamical} {Evolution} of {Boson} {Stars}: {Perturbing} the
  {Ground} {State}},
        date={1990-07},
     journal={Phys. Rev. D},
      volume={42},
       pages={384\ndash 403},
      eprint={http://link.aps.org/doi/10.1103/PhysRevD.42.384},
}

\bib{Sharma08}{article}{
      author={Sharma, R.},
      author={Karmakar, S.},
      author={Mukherjee, S.},
       title={Boson {Star} and {Dark} {Matter}},
        date={2008-12},
     journal={[arXiv:0812.3470 [gr-qc]]},
      eprint={http://arxiv.org/abs/0812.3470},
}

\bib{Sin94}{article}{
      author={Sin, S.-J.},
       title={Late-{Time} {Phase} {Transition} and the {Galactic} {Halo} as a
  {Bose} {Liquid}},
        date={1994-09},
     journal={Phys. Rev. D},
      volume={50},
       pages={3650\ndash 3654},
      eprint={http://link.aps.org/doi/10.1103/PhysRevD.50.3650},
}

\end{biblist}
\end{bibdiv}

\end{document}